\documentclass[12pt]{article}
\newcommand{\journal}[4]{{#1~}{#2}\,(#3)\,#4.}

\newcommand{\ijmp}{\journal {Int. J. Mod. Phys.}}

\newcommand{\jmp}{\journal {J. Math. Phys.}}

\newcommand{\cmp}{\journal {Comm. Math. Phys.}}

\newcommand{\np}{\journal {Nucl. Phys.}}
\newcommand{\pl}{\journal {Phys. Lett.}}
\newcommand{\mpl}{\journal {Mod. Phys. Lett.}}

\newcommand{\annp}{\journal {Ann. Phys. }}
\newcommand{\euro}{\journal {Europhys. Lett.}}

\newcommand{\PR}{\journal{Phys. Rev.}}

\makeatletter \@addtoreset{equation}{section} \makeatother

\def\text#1{\mbox{#1}}
\def\equ#1{(\ref{#1})}

\def\be#1{\begin{equation}\label{#1}}
\def\ee{\end{equation}}
\def\equ#1{(\ref{#1})}
\def\Tr{\mbox{Tr}}

\def\hodge#1{{ }^{*}\!#1}
\def\cal{\mathcal}
\def\pa{\partial}

\begin{document}

\begin{titlepage}
\title{{Field redefinitions and massive BF models in arbitrary space-time dimensions}}
 \author{R. R. Landim\thanks{Electronic address: renan@fisica.ufc.br}\\
Universidade Federal do Cear\'{a} - Departamento de
F\'{\i}sica \\ C.P. 6030, 60470-455 Fortaleza-Ce, Brazil}


\end{titlepage}
 \maketitle
  \begin{abstract}
 We show that the topological massive BF theories can be written
  as a pure BF term through  field redefinitions. The fields are rewritten as power expansion
  series in the inverse of the mass parameter $m$. We also give a cohomological justification
  of this expansion through  BRST framework.
  In this approach the BF term can be seen as a topological generator for massive BF theories.
  
\end{abstract}
\vspace{1.0cm}


\vspace{0.3cm}


\section{Introduction}

After the work of Deser, Jackiw  and Templeton \cite{jackiw}, topological massive theories
have been object of  continuous source of investigation from both  mathematical and physical point of
views. Such theories represent an alternative to give mass to gauge fields
without the Higgs mechanism. In three dimensions they can be formulated with the
Chern-Simons term in an abelian and non-abelian version \cite{jackiw}. Its
generalization to any dimension is possible via  the  topological BF term, where
B is a $(D-2)$-form gauge field and F is the field  strength of the usual vector
gauge field \cite{lah,hwang,lan}.  It is also worthwhile to mention that other
topological  terms have been used to construct topological massive
theories~\cite{non-chern,oda}.

A lot of work has been done in order to have a better understanding of topological massive theories,
leading
to interesting and promising results. An important result was obtained a few years ago
by Giavarini and collaborators about the finiteness of  topological massive Yang-Mills theory in
three dimensions, with a careful analysis of higher loops of the Feynman integrals \cite{gi1}.
More recently, Lemes {\it et al} \cite{sor1,sor2}, showed that the topological three dimensional massive
Yang-Mills theory can be cast in the form of a pure Chern-Simons action through a non-linear
but local field redefinition,
\be{cher-red}
S_{YM}(A)+S_{CS}(A)=S_{CS}(\hat{A}),
\ee
where
\begin{eqnarray}
&&S_{YM}=\frac{1}{4m}\Tr\int d^3x~F_{\mu\nu}F^{\mu\nu},\\
&&S_{CS}(A)=\frac{1}{2}\int d^3x~\varepsilon^{\mu\nu\rho}
\left(A_\mu\pa_\nu A_\rho+\frac{2}{3}g A_\mu A_\nu A_\rho\right),
\end{eqnarray}
and
\be{A-red}
\hat{A}_\mu=A_\mu+\sum_{n=1}^\infty \frac{1}{m^n}\vartheta_\mu^n(D,F).
\ee
As shown by the authors \cite{sor2}, the coefficients $\vartheta_\mu^n(D,F)$ are
local and covariant and depend only on the field strength $F_{\mu\nu}$ and
the covariant derivative $D_\mu=\pa_\mu+g[A_\mu,~]$. This property, also valid for the
abelian Chern-Simons theory, was  used in bosonization in three dimensions~\cite{sor3}.

In four dimensions, the contribution from one loop diagrams to the effective
action in theories with four fermions
reproduces the topological  massive BF theory \cite{leblanc}. The topological
massive BF theory also appears in D-dimensional
bosonization of massive Thirring model \cite{banerj}.

The BF term is not important only in the construction of topological massive theories in any
dimension, but also in other branches of physics. They  are related to the Ray-Singer
torsion~\cite{scw} and the intersection number of manifolds in any dimension \cite{blau}.
Also they provide an example of ultraviolet finite field theories \cite{uf}.  A few years ago a
generalization of anyons to (3+1) dimensions, making use of the BF term, was proposed in \cite{ber}.
    More recently, Smailagic and Spallucci, have studied
the dualization of abelian \cite{spa} and nonabelian \cite{spa1} BF models
of arbitrary $p-$forms to a Stueckelberg-like massive gauge invariant
theories.

The aim of this work is to show that the topological massive BF theories in
$D$-dimensions can also be written as a pure BF term through  field 
redefinitions. This feature is not exclusive to the topological three
dimensional massive Yang-Mills theory, but seems to be a property of all
topological massive theories in any dimension.

The paper is organized as follows. In section 2 we analyse the field redefinitions in the abelian topological
BF model, where  a cohomological justification through the BRST framework is established.
The section  3 is devoted to a brief discussion about
the field redefinitions of the non-abelian topological massive BF model.
Further possible applications will be
discussed.

\section{The abelian case}
Let us consider the abelian massive BF model in $D$-dimensions, described by
 a real-valued $(D-2)-$form field $B$ and a
real-valued $1-$form field $A$, both with canonical dimension $(D-2)/2$, defined 
in a D-dimensional space-time manifold ${\cal M}_D$ with metric $g_{\mu \nu
}=\mbox{diag}(-++\cdots +++)$, with the action given by \begin{equation}
S_0=\int_{{\cal M}_D}\left( \frac 12H\wedge ^{*}\!\!H+mB\wedge
F-\frac 12F\wedge ^{*}\!\!F\right) ,  \label{act}
\end{equation}
where $F=dA$ and $H=dB$ are the field strengths of $A$
and $B$ respectively, $m$ is a mass parameter, $d=dx^\mu (\partial /\partial x^\mu )$ is the exterior
derivative and $*$ is the Hodge star operator\footnote{We shall use the form
representation of the fields in order to simplify our analysis.}. The
 adjoint operator acting in a $p-$form is defined as $d^{\dagger
}=(-1)^{Dp+D}*d*$\cite{nakahara}.

The action~\equ{act}  is invariant under the gauge
transformations

\begin{eqnarray}
&\delta A=d\theta,  \nonumber \\
&\delta B=d\Omega,  \label{trans}
\end{eqnarray}
where $\theta $ is a $0-$form and $\Omega $ is a $(D-3)-$form.

We will show that the action \equ{act} is indeed a pure BF term through a local and linear
field redefinition of the fields $A$ and $B$, namely
 \be{bf-red}
S_0=\int_{{\cal M}_D}\left( \frac 12H\wedge ^{*}\!\!H+mB\wedge
F-\frac 12F\wedge ^{*}\!\!F\right)=\int_{{\cal M}_D}m\hat{B}\wedge
\hat{F} ,
\ee
where $\hat{F}=d\hat{A}$, and
\begin{eqnarray}
&&\hat{B}=B+\sum_{i=1}^\infty \frac{B_i}{m^i},\label{red1}\\
&&\hat{A}=A+ \sum_{i=1}^\infty \frac{A_i}{m^i}. \label{red2}
\end{eqnarray}
The terms $B_i$ and $A_i$ are $(D-2)$-form and 1-form respectively, constructed only
with the field strength $H$ and $F$, the Hodge operator and the exterior derivative.  The power
in the inverse of mass parameter $m$ in \equ{red1} and \equ{red2}  gives the correct mass dimension for
$\hat{B}$ and $\hat{A}$.

The properties given by equations \equ{bf-red}, \equ{red1} and \equ{red2} seem to be  fundamental
 not only  for the Chern-Simons theory but for all topological massive theories.

In order to provide an explicit form for the terms $B_i$ and $A_i$, we insert the redefined field
given by eqs. \equ{red1} and \equ{red2} into the eq.\equ{bf-red} and
identify the terms with the same power in $1/m$.  We find the following expressions,
 \begin{eqnarray}
&&B_1=-\frac{1}{2}\hodge{F},\nonumber\\
&&A_1= \frac{1}{2}(-1)^{D-1}\hodge{H},\label{first}
\end{eqnarray}
and the integral consistency condition
\be{cons}
\int_{{\cal M}_D}\left(B_{k+1}\wedge dA+
B \wedge dA_{k+1}+\sum_{i=1}^k B_i dA_{k+i-1}\right)=0,\quad \mbox{for}\quad k \ge 1.
\ee
 Contrary to the Chern-Simons case in three dimensions~\cite{sor1,sor2}, we have many solutions to
eq. \equ{cons}. This is due to the fact that we start with two fields, $A$ and $B$, in the
original theory. We give one solution of \equ{cons} explicitly
\begin{eqnarray}
&&A_{2j}=0,\\
&&A_{2j+1}=\frac{1}{2^{j+2}}(-1)^{(j+1)D+1}(\hodge{d}\!~)^{2j}~\hodge{H},\label{a2j1}\\
&&B_{2j}=\frac{1}{2} \hodge{d} A_{2j-1},\\
&&B_{2j+1}=0, \quad j\ge 1.
\end{eqnarray}
The term $A_1$ in $D=4$, which corresponds to $j=0$ in eq.\equ{a2j1}, was used 
in \cite{amorim} to couple fermions to Kalb-Ramond field. As shown by the
authors, this coupling leads an anomalous axial current.

  Let us underline that the equation ~\equ{bf-red} has to be understood here in
pure classical terms. The presence of the expansion parameter $1/m$ in
Eqs.\equ{red1} and \equ{red2}, will introduce an infinite number of power
counting nonrernomalizable interactions and a more careful  quantum
analisys should be done. In the present work we are interested only classical
aspects of the field redefinitions.

As in the Chern-Simons case~\cite{sor1,sor2}, there is an analog cohomogical
justification to eq. \equ{bf-red}. This justification is based on
~\cite{sor1,sor2}  and is quite formal.  Let $\widetilde{B}$ and $\widetilde{A}$
be the anti-fields of $B$ and $A$. Obviously we need a pyramid of ghosts to take
into account of the reducibility of the gauge transformation of $B$, but not
necessary for our purposes. Since we are interested only in the classical
aspects of \equ{bf-red}, there is no necessity of a gauge fixing term. The
dependence of $\widetilde{B}$ and $\widetilde{A}$ is given by
\cite{piguet,henneaux}
\be{anti-dep}
S(\widetilde{A},\widetilde{B})=\int_ {{\cal M}_D} \widetilde{A}\wedge\hodge{d}c+
\widetilde{B} \wedge\hodge{d}\eta,
 \ee
 where $c$ and $\eta$ are the ghosts for the gauge transformations \equ{trans},
namely   \begin{eqnarray}
&sA=dc,  \label{brsa}\\
&sB=d\eta.  \label{brsB}
\end{eqnarray}
Following the standard procedure \cite{piguet,henneaux}, the BRST transformation
of the anti-fields $\widetilde{B}$ and $\widetilde{A}$ are
\begin{eqnarray}
&&s\widetilde{B}=d^{\dagger}H-m~\hodge{F},\label{sbtil}\\
&&s\widetilde{A}=-d^{\dagger}F+(-1)^Dm~\hodge{H}\label{satil}.
\end{eqnarray}
The equations \equ{sbtil} and \equ{satil} can be writing  in a suitable form
\begin{eqnarray}
&&H=\frac{1}{m^2}(-1)^D dd^{\dagger}H+
\frac{1}{m}s\left(\frac{1}{m}(-1)^D d\widetilde{B}+ \hodge{\widetilde{A}}\right),\label{H} \\
&&F=\frac{1}{m^2}dd^{\dagger}F-
\frac{1}{m}s\left(\frac{1}{m}d\widetilde{A}+ (-1)^D~\hodge{\widetilde{B}}\right).\label{F}
\end{eqnarray}
The equations \equ{H} and \equ{F} are recursive formulas for $H$ and $F$ respectively. This
allows us to write  $H$ and $F$ as a BRST trivial, namely
\begin{eqnarray}
H=s\left(\sum_{i=1}^\infty \frac{H_i}{m^i}\right),\label{sH}\\
F=s\left(\sum_{i=1}^\infty \frac{F_i}{m^i}\right).\label{sF}
\end{eqnarray}
This property enables us to express the kinetic terms of \equ{act}  as a pure BRST variation.
It is well known, from the BRST algebraic framework, that terms of the action which are BRST trivial
correspond to field redefinitions.  Therefore,
the expressions \equ{bf-red}, \equ{red1} and \equ{red2} are consequences of the BRST
triviality of   $F$ and $H$.  Let us underline that the triviality
of  $F$ and $H$  is due to the presence of the BF term in the action \equ{act}.  As we can
see from \equ{sbtil} and \equ{satil}, the absence of the BF term  spoils the possibility
of writing $F$ and $H$ as a recursive formula.

 \section{The non-abelian case} In this section we consider the
topologically massive non-abelian BF model in D-dimensions. We will use the same
convention of \cite{lan}.  The gauge fields are written as $A=A^aT^a$,
$B=B^aT^a$, where $T^a$ are generators of a Lie algebra ${\cal G}$ of a
semi-simple Lie group $G$.  The action reads \be{non-abel} S=\Tr\int_{{\cal
M}_D}\left( \frac 12H\wedge ^{*}\!\!H+mB\wedge F-\frac 12F\wedge
^{*}\!\!F\right),  \ee  where $F=dA+A\wedge A$ and $H=DB+[F,V]$.  $D=d+[A, ~]$
is the covariant derivative and   $V=V^aT^a$ is an auxiliary $(D-3)$-form Lie
algebra valued, with canonical dimension $(D-4)/2$.  This auxiliary field is
necessary in order to implement the gauge invariance of the model and to
eliminate a constraint that appears in the equations of motion \cite{lan}.  The
action \equ{non-abel} is invariant under the gauge  transformations\footnote{%
The commutator between two Lie algebra valued forms $P$ and $Q$ is defined
as $[P,Q]=g\left(P\wedge Q-(-1)^{d(P)d(Q)}Q\wedge P\right)$, where $d(X)$ is the
form degree of $X$ and $g$ is a parameter with mass dimension
${(4-D)}/{2}$.}  \begin{eqnarray}
&& \delta B=D\Omega+[B,\theta],\\
 &&\delta A =D\theta,\\
 &&\delta V=-\Omega+[V,\theta].
 \end{eqnarray}
We shall show that the action \equ{non-abel} can be written as a pure BF term, namely

\be{pure-bf}
S=\Tr\int_{{\cal M}_D}\left( \frac 12H\wedge ^{*}\!\!H+mB\wedge
F-\frac 12F\wedge ^{*}\!\!F\right)=\Tr\int_{{\cal M}_D}m\hat{B}\wedge
\hat{F} ,
\ee
where $\hat{F}=d\hat{A}+\hat{A}\wedge \hat{A}$, and
\begin{eqnarray}
&&\hat{B}=B+\sum_{i=1}^\infty \frac{1}{m^i}\left(B_i-[A_i,V]\right),\label{red3}\\
&&\hat{A}=A+ \sum_{i=1}^\infty \frac{A_i}{m^i}. \label{red4}
\end{eqnarray}
 The terms $A_i$ and $B_i$ are 1-form and  $(D-2)$-form respectively, constructed with $H$, $F$,
 the covariant derivative $D$ and the hodge operator. They are
 non-linear due to the non-abelian character of the model. Proceeding as in the  abelian case,
 we show bellow some terms of the expansions \equ{red3} and \equ{red4},
 \begin{eqnarray}
 &&A_1= \frac{1}{2}(-1)^{D-1}~\hodge{H}, \nonumber\\
 &&A_2=\frac{1}{8}(-1)^{D-2}~\hodge{}\left[B+DV,\hodge{H}\right],\nonumber\\
 &&A_3=-\frac{1}{2}\hodge{}\left[B+DV, A_2\right]-\frac{1}{8}~\hodge{D}\hodge{D}\hodge{H}, \\
 &&B_1=-\frac{1}{2}\hodge{F},\nonumber\\
 &&B_2=\frac{1}{2}\hodge{DA_1},\nonumber\\
 &&B_3=\frac{1}{2}\hodge{DA_2}+\frac{1}{4}~\hodge{\left[A_1,A_1\right] }. \nonumber
 \end{eqnarray}
 Let us underline that the terms $A_i$ and $B_i$ transforms covariantly, i.e,
 \begin{eqnarray}
 &&\delta A_i=\left[A_i,\theta\right],\\
 && \delta B_i=\left[B_i,\theta\right].
 \end{eqnarray}
 This property can be easily obtained by the gauge invariance of both sides of \equ{pure-bf}. Note that the
 field $V$ do not need to be redefined. This is due to the fact that $V$ is an auxiliary field. The covariant transformation of
 $A_i$ and $B_i$ leads to   the following gauge transformation for $\hat{A}$ and $\hat{B}$,
\begin{eqnarray}
&& \delta \hat{B}=\hat{D}\Omega+[\hat{B},\theta],\label{dhatb}\\
 &&\delta \hat{A} =\hat{D}\theta,
 \end{eqnarray}
 where $\hat{D}$ is the redefined covariant derivative,
 \be{red-der}
 \hat{D}=d+[\hat{A},~].
  \ee
 Just as in the abelian case, the field redefinition is a consequence of the BRST
 triviality of the fields strengths $H$ and $F$, due
 to the presence of the BF term.  In order to justify this, we  follow the
 same arguments used for the Chern-Simons case \cite{sor2}. The full BRST differential $s$, in this case,
 has non-linear dependence on the fields and
 anti-fields.  We can filter the linear part, say $s_0$, from the complete BRST operator $s$.
 There is a theorem of the BRST cohomology~\cite{piguet,henneaux}, which states
that the cohomology of the full  BRST operator $s$ is isomorphic to the
cohomology of the corresponding linear operator $s_0$.  The linear operator
$s_0$ is just the BRST operator of the abelian case discussed in section 2.  As
proved in section 2, the abelian fields strengths $H$ and $F$ are BRST trivial
in a formal  power series expansion in $1/m$. Hence, making the use of the above
theorem,  the non-abelian $H$ and $F$ can be  written as a pure $s$ variation  in
a formal power series expansion in the  parameter $1/m$.  \section{Conclusion}
 In this paper we have shown that  both abelian and non-abelian topological massive
 BF theories can be written as a pure BF term through  field redefinitions. This result, valid for
 any dimension, represents a generalization of the results obtained in  \cite{sor1,sor2} for topological three dimensional massive
Yang-Mills theory. Let us underline that this property seems to be
 valid for all topological massive theories in any dimension. In this framework
  the BF term can be seen as a topological generator for massive BF theories.

  Let us comment about a possible application of our result.  As showed in \cite{sor3}
  the three dimensional abelian fermionic
  determinant   is a pure  Chern-Simons term through local and non-local field redefinitions.
  This suggest that abelian fermionic determinant in $D$ dimension could be
  written in a suitable way as a pure BF term through field
  redefinitions. Also our result can be applied  to the study
  of the finiteness of the  topologically massive non-abelian BF models.

  \vskip2cm
  \noindent
  {\large\bf Acknowledgments}
  \vskip0.5cm

  We wish to thank to J. R. Goncalves for reading the manuscript. The Conselho Nacional de
   Desenvolvimento Cient\'\i fico e tecnol\'ogico-CNPq is gratefully acknowledged for financial support.

   This paper is dedicated to the memory of my father.

\end{document}